\documentclass[showpacs,preprintnumbers,amsmath,amssymb]{revtex4}

\usepackage{graphicx}
\usepackage{dcolumn}
\usepackage{bm}
\usepackage{epsfig}
\usepackage{amsfonts}

\begin{document}

\title{Central limit behavior of deterministic dynamical systems}
\author{Ugur Tirnakli}
\affiliation{Department of Physics, Faculty of Science, Ege
University, 35100 Izmir, Turkey}

\author{Christian Beck}
\affiliation{School of Mathematical Sciences, Queen Mary,
University of London, Mile End Road, London E1 4NS, UK}

\author{Constantino Tsallis}
\affiliation{Centro Brasileiro de Pesquisas F\'\i sicas,
R. Dr.  Xavier Sigaud 150, 22290-180 Rio de Janeiro, RJ, Brazil}

\date{\today}

\begin{abstract}
We investigate the probability density of rescaled sums of
iterates of deterministic dynamical systems, a problem relevant
for many complex physical systems consisting of dependent random
variables. A Central Limit Theorem (CLT) is only valid if the
dynamical system under consideration is sufficiently mixing.
For the fully developed logistic map and a cubic map we
analytically calculate the leading-order corrections to the CLT
if only a finite number of iterates is added and rescaled, and
find excellent agreement with numerical experiments. At the
critical point of period doubling accumulation, a CLT is not
valid anymore due to strong temporal correlations between the
iterates. Nevertheless, we provide numerical evidence that in
this case the probability density converges to a $q$-Gaussian,
thus leading to a power-law generalization of the CLT. The above
behavior is universal and independent of the order of the maximum
of the map considered, i.e.\ relevant for large classes of
critical dynamical systems.
\end{abstract}

\maketitle

The Central Limit Theorem (CLT) is an extremely important concept
in probability theory and it also lies at the heart of
statistical physics \cite{vKa,khinchin}. It basically says that the sum of
$N$ independent identically distributed (IID) random variables, rescaled with a factor
$1/\sqrt{N}$, has a Gaussian distribution in the limit $N\to
\infty$. The CLT plays a crucial role in explaining why many
stochastic processes that are of relevance in physics, chemistry,
biology, economics, etc. are Gaussian, provided they consist of a
sum of many independent or nearly independent contributions. CLTs
are also of fundamental importance to `derive' statistical
mechanics from first principles: If a CLT is valid for the
driving forces in a many-body system, it is easy to proceed to
the formalism of statistical mechanics via the Langevin and Fokker-Planck approaches.

What is less known in the physics community but well-known in the
mathematics community is the fact that there are also CLTs for
the iterates of {\em deterministic} dynamical systems. The
iterates of a deterministic dynamical system can never be
completely independent, since they are generated by a
deterministic algorithm. However, if the assumption of
IID is replaced by the weaker property that the
dynamical system is sufficiently strongly mixing, then various
versions of CLTs can be proved for deterministic dynamical systems
\cite{bill, kac, chernov, keller, beck, roep,
michael}. It should be kept
in mind that the mixing property just means asymptotic statistical
independence for large time differences.

In this letter we investigate in detail the central limit
behavior of deterministic systems. We are interested in two
questions that are of fundamental importance for the foundations
of statistical mechanics: a) Suppose a CLT is valid for a
deterministic dynamical system for $N\to \infty$, what are the
leading-order corrections to the CLT for large but finite $N$? b)
Suppose the dynamical system does not satisfy a CLT because it is
not sufficiently mixing, what are typical probability
distributions that one obtains for these types of systems in the
limit $N \to \infty$?

The above questions are very relevant to understand the physics
of complex systems in general. First of all, any physical system
always consists of a finite number $N$ of constituents rather
than an infinite one. Hence finite-$N$ corrections to the CLT can
be potentially important for small systems.  Secondly, the
dynamics of critical systems usually exhibits strong
correlations. These imply that an ordinary CLT cannot be valid.
In such cases it is important to know what type of distributions
replace the usual Gaussian limit distributions.

In full generality, the above two problems are very difficult to
deal with. Hence there is the need to start with simple model
systems where some statements can be rigorously proved. Our main
example in the following is the logistic map. For the fully
developed chaotic state of this map, a CLT has been proved
\cite{bill}. We will explicitly calculate leading-order
corrections to the Gaussian limit case if a finite number $N$ of
iterates is added and rescaled with $1/\sqrt{N}$. Moreover, we
will provide evidence that at the critical point of period
doubling accumulation, where a CLT is not valid due to strong
correlations between the iterates, a suitably rescaled sum of
iterates appears to generate distributions with power law tails,
which are well approximated by $q$-Gaussians. These distributions
are known to play an important role in generalized versions of
statistical mechanics \cite{tsa1,tsa2,tsa3}. Although our results
are derived for the special example of the logistic map, we will
show that they are universal, i.e., applicable to entire classes
of deterministic dynamical systems.

To start with, let us consider a $d$-dimensional mapping of the
form
\begin{equation}
x_{i+1}=T (x_i)
\end{equation}
on some $d$-dimensional phase space $X$. If $T$ is sufficiently
strongly mixing (see \cite{bill, roep, chernov, michael} for technical
details), one can prove the existence of a CLT. This means the
probability distribution of
\begin{equation}
y:= \frac{1}{\sqrt{N}} \sum_{i=1}^N f(x_i),
\end{equation}
becomes Gaussian for $N \to \infty$, regarding the initial value
$x_1$ as a random variable. Here $f: {\cal R}^d \to {\cal R}^k $
is a suitable smooth function with vanishing average which
projects from the $d$-dimensional phase space to a $k$-dimensional
subspace. If $d=k=1$, the variance $\sigma^2$ of this Gaussian is
given by \cite{roep}
\begin{equation}
\sigma^2 =\langle f(x_0)^2 \rangle +2 \sum_{i=1}^\infty \langle
f(x_0) f(x_i) \rangle . \label{sigma}
\end{equation}
Here $\langle \cdots \rangle$ denotes
 an expectation formed with
the natural invariant density of the map $T$.

\begin{figure}
\epsfig{file=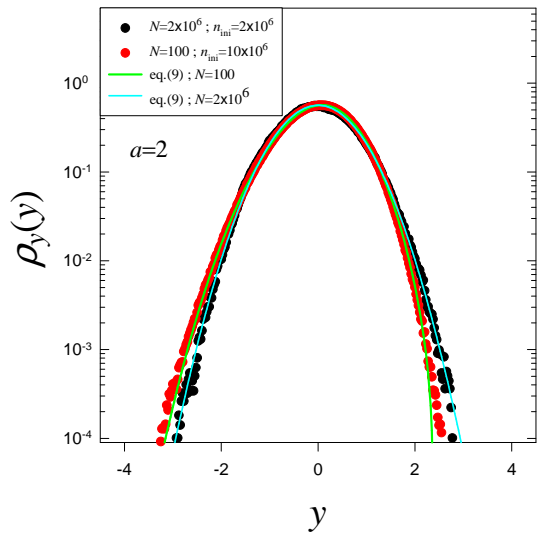, width=7cm, height=7cm}
\caption{(Color online) Probability density of rescaled sums of iterates of the
logistic map with $a=2$ as given by eq.~(\ref{Y}), $N=2\times 10^6$
and $N=100$. The number of initial values contributing to the
histogram is $n_{ini}=2 \times 10^6$, respectively
$n_{ini}=10^7$. The solid lines correspond to eq.~(\ref{here}).}
\end{figure}

As an example to illustrate these general results, let us consider
the logistic map
\begin{equation}
x_{i+1}=T(x_i)=1-a x_i^2 \label{logi}
\end{equation}
on the interval $X=[-1,1]$. For $a=2$ the system is
(semi)conjugated to a Bernoulli shift and strongly mixing. The
natural invariant density is
\begin{equation}
\rho_x(x) =\frac{1}{\pi \sqrt{1-x^2}}. \label{inva}
\end{equation}
Ergodic averages of arbitrary observables $A$ are given by
\begin{equation}
\langle A(x) \rangle =\int_{-1}^1 \rho (x) A(x) dx.
\end{equation}
The average $\langle x \rangle$ vanishes. For the correlation
function one has
\begin{equation}
\langle x_{i_1} x_{i_2} \rangle =\frac{1}{2} \delta_{i_1, i_2}.
\label{co}
\end{equation}
Due to the strong mixing property, the conditions for the
validity of a CLT are satisfied for $a=2$. This means the distribution of
the quantity
\begin{equation}
y:= \frac{1}{\sqrt{N}} \sum_{i=1}^N (x_i -\langle x \rangle)
\label{Y}
\end{equation}
becomes Gaussian for $N\to \infty$, regarding the initial value
$x_1$ as a random variable with a smooth probability distribution.
For the variance of this Gaussian we obtain from eq.~(\ref{sigma})
and (\ref{co}) the value $\sigma^2=\frac{1}{2}$ (choosing
$f(x)=x$). The above CLT result is highly nontrivial, since there
are complicated higher-order correlations between the iterates of
the logistic map for $a=2$ (see \cite{nonli} for details). This
means the ordinary CLT, which is only valid for {\em independent}
$x_i$ cannot be directly applied, an extension of the CLT for
mixing systems is necessary \cite{bill}.

For physical and practical applications, the number $N$ is always
finite, hence it is important to know what the finite-$N$
corrections are for a given dynamical system. This problem can be
solved for our example, the map $T(x)=1-2x^2$, by applying the
general graph-theoretical methods developed in
\cite{nonli,hilgers}. Our final result is that for finite but
large $N$ the probability density of $y$ is given by
\begin{equation}
\rho_y (y)= \frac{1}{\sqrt{\pi}}e^{-y^2} \Big[
1+\frac{1}{\sqrt{N}}y \left( \frac{3}{2} -y^2 \right) +O\left(
\frac{1}{N} \right) \Big] . \label{here}
\end{equation}
This result is in excellent agreement with numerical experiments
(Fig.~1).

\begin{figure}
\epsfig{file=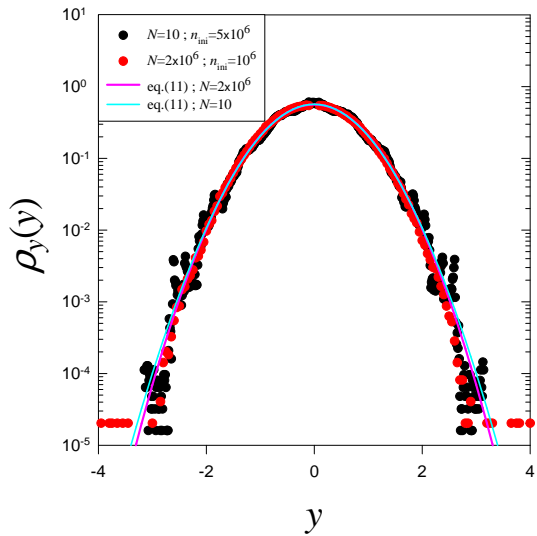, width=7cm, height=7cm}
\caption{(Color online) Probability density of rescaled sums of iterates of the
cubic map (\ref{cubic}) for $N=10^7$ and $N=10$. The number of
initial values is $n_{ini}=10^6$, respectively $n_{ini}=5 \times
10^6$. The solid lines correspond to eq.~(\ref{there}).}
\end{figure}

We should mention at this point that the finite-$N$ corrections
are non-universal, i.e. different mappings have different
finite-$N$ corrections. For example, by applying the techniques of
\cite{nonli,hilgers} to the cubic map
\begin{equation}
x_{i+1}=4x_i^3-3x_i,\label{cubic}
\end{equation}
which has {\em the same} invariant density (\ref{inva}) as the
logistic map with $a=2$ but {\em different} higher-order
correlations, we obtain in leading order
\begin{equation}
\rho_y (y) =\frac{1}{\sqrt{\pi}} e^{-y^2} \Big[ 1+ \frac{1}{N}
\left( \frac{1}{12} y^4-\frac{1}{4} y^2 +\frac{1}{16} \right)
\Big] . \label{there}
\end{equation}
Note that in this case the leading-order corrections are of order
$1/N$, rather than of order $1/\sqrt{N}$. Again our analytical
result is in good agreement with the numerics, see Fig.~2.

Apparently the finite-$N$ corrections to the asymptotic Gaussian
behavior of dynamical systems satisfying a CLT are non-universal
and can be used to obtain more information on the underlying
deterministic dynamics. This is certainly important from a
general physical point of view if the dynamics underlying a CLT
is a priori unknown.

We may also look at distributions of the variable $y$ obtained for
`typical' parameter values in the chaotic regime of the logistic
map, such as $a=1.7, 1.8, 1.9$. A CLT has not been rigorously
proved in this case, but nevertheless we again observe Gaussian
limit behavior. This is shown in Fig.~3. It is a well-known
fact that for $a<2$ the
average $\langle x \rangle$ does not vanish anymore.
The Gaussians observed in Fig.~3 have smaller variance $\sigma^2$ as compared
to the case $a=2$. This can be understood in a quantitative way
from eq.~(\ref{sigma}).
The fits in Fig.~3 show
Gaussians with variance parameter $\sigma^2$ directly determined
from eq.~(\ref{sigma}) using $f(x)=x- \langle x \rangle$ (the averages
$\langle \dots \rangle$ are calculated as time averages). Note that for
$a=1.7,1.8,1.9$ the correlation function is not
$\delta$-correlated anymore.

\begin{figure}
\epsfig{file=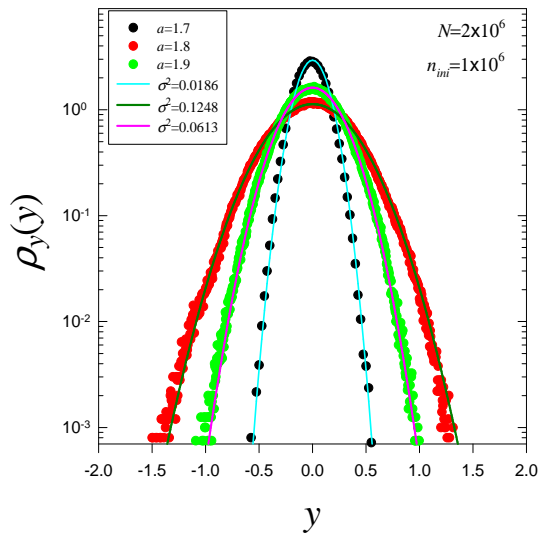, width=7cm, height=7cm}
\caption{(Color online) Probability density of rescaled sums of iterates of the
logistic map as given by eq.~(\ref{Y}) for $a=1.7,1.8,1.9$
and $N=2\times 10^6$, $n_{ini}=10^6$. The solid lines show Gaussians
$e^{-y^2/(2\sigma^2)}/\sqrt{2\pi \sigma^2}$
with variance parameter $\sigma^2$ determined from eq.~(\ref{sigma}).
}
\end{figure}

Next, let us investigate the behavior of deterministic dynamical
systems where the conditions for a CLT are {\em not} satisfied.
As a particularly interesting example, we choose the logistic map
at the accumulation point of period doublings (i.e., at the edge of chaos). The corresponding
parameter value is denoted by $a_c$. For $a=a_c$ the logistic map
is not sufficiently strongly mixing and a CLT is not valid. Still
one can ask the question if there is a universal limit
distribution for suitably rescaled sums of iterates. We
investigated this question numerically and looked at rescaled sums
of the form
\begin{equation}
y=N^\gamma \sum_{i=1}^N (x_i -\langle x\rangle). \label{YY}
\end{equation}
Eq.~(\ref{YY}) is a generalization of eq.~(\ref{Y}) with a more
general rescaling exponent $\gamma$. The
notation $\langle \cdots \rangle$ means an average over a large
number $N$ of iterations {\em and} a large number $n_{ini}$ of
randomly chosen initial values $x_1^{(j)}$. Numerically we
calculate
\begin{equation}
\langle x \rangle = \frac{1}{n_{ini}} \frac{1}{N}
\sum_{j=1}^{n_{ini}} \sum_{i=1}^N x_i^{(j)}.
\end{equation}
Due to the fact that the system is not necessarily ergodic
anymore, the average over initial conditions is an important
ingredient.

\begin{figure}
\epsfig{file=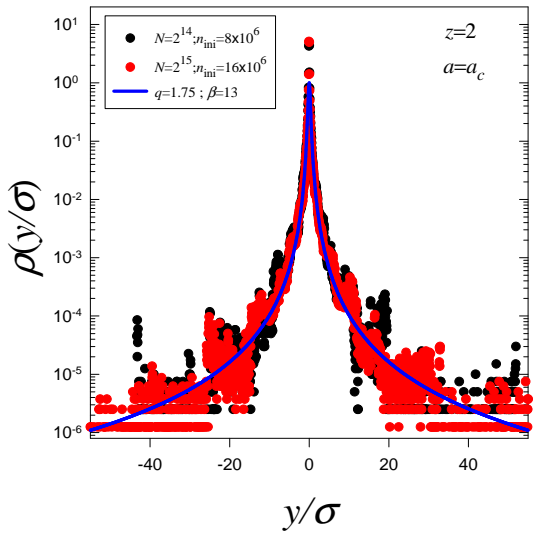, width=7cm, height=7cm}
\caption{(Color online) Probability density of the quantity $y/\sigma$ at the critical
point $a_c$ for $z=2$, $N=2^{14}$ and $N=2^{15}$.}
\end{figure}

In our numerical experiment, a large number of initial values
$x_1$ were randomly chosen and the corresponding values of the
sum $Y$ were plotted in a histogram. We then looked for a suitable
rescaling exponent $\gamma$ where we have data collapse, i.e., the
same shape of the probability distributions $\rho_y (y)$ of $y$ if
$N$ is increased. Numerically we observe that, at $a=a_c$, this
value is given by $\gamma =1.5$. The density of $y$ is not given
by a Gaussian, as one would expect if a CLT is valid, but well
fitted by a $q$-Gaussian, i.e. a distribution of the form
\begin{equation}
\rho_y (y) \sim e_q^{-\beta y^2} :=
\frac{1}{(1+\beta (q-1) y^2)^{\frac{1}{q-1}}},
\end{equation}
where $q$ and $\beta$ are suitable parameters.
We observe $q = 1.75 \pm 0.03$ (see Fig.~4). The rescaling factor
$N^\gamma$ in eq.~(\ref{YY}) can be absorbed by simply
calculating the variance $\sigma^2$ of the unrescaled sum
$y:=\sum_{i=1}^N (x_i-\langle x\rangle)$ for a given $N$ and then
plotting a histogram of $y/\sigma$.

Interesting enough, the values of $q$ and $\gamma$ that we
observe in our numerical experiments are independent of the order
of the maximum of the map considered. Indeed, if instead of the logistic
map (\ref{logi}) we iterate the more general map
\begin{equation}
x_{i+1} =1-a |x_i|^z,
\end{equation}
then at the critical point $a_c(z)$ of period doubling accumulation
the results are basically unchanged (see Fig.~5).

Since at the critical point the dynamics of many different maps
converges to a dynamics given by the universal Feigenbaum fixed
point function \cite{fei}, our results for the asymptotic
probability distribution of $Y$ are universal: Entire classes of
critical quadratic maps will generate the same $q$-Gaussian limit
distribution. Our result is even more universal since there seems
to be no dependence on the order $z$ of the maximum of the map.
Thus we expect the $q$-Gaussian limit distribution with $q \approx
1.75$ to be relevant for many different dynamical systems at the
critical point.

The independence of $z$ has also been established in a different
context: In
\cite{robledo2004} it is shown that, {\it for all values of $z$}, the
relevant fixed point map describing period doubling bifurcations (tangent bifurcations) is
a specific $q$-exponential with $q=3$ ($q=2$).

\begin{figure}
\epsfig{file=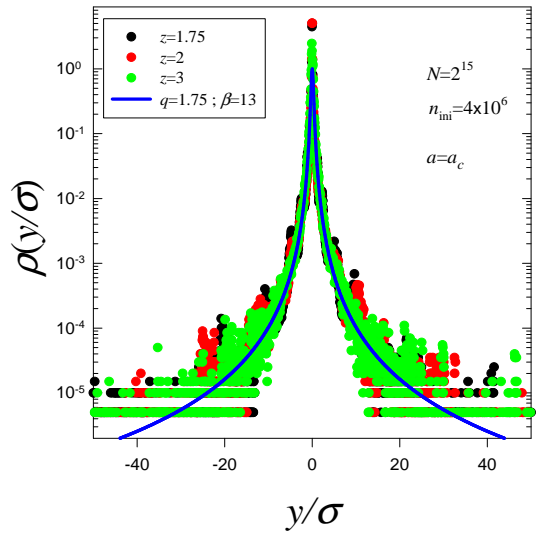, width=7cm, height=7cm}
\caption{(Color online) Probability density of the quantity $y/\sigma$ at the critical
point $a_c$ for $z=1.75,2,3$.}
\end{figure}

The generalized central limit behavior of critical dynamical
systems observed in this paper may be of relevance for more
general classes of critical systems in physics as well. For
example, Caruso et al. \cite{rapi} observe that the probability
distribution of energy differences of subsequent earthquakes in
the World Catalog and in Northern California is well fitted by a
$q$-Gaussian with $q\approx 1.75$. Their model for this is based on
self-organized criticality and the OFC model. We note that
$q$-Gaussians with $q\approx 1.75$ arise naturally if the
corresponding random variable consist of a sum of strongly
correlated contributions as generated by critical dynamical
systems.
A somewhat similar result has also been recently
observed for Brazilian financial
data: A $q$-Gaussian with $q\simeq 1.75$ fits
histograms of stock market index changes
for a considerable range of time delays (see Fig. 6 of \cite{cortines}).

To conclude, in this paper we have  explicitly calculated
finite-$N$-corrections to the CLT for some examples of strongly
mixing dynamical systems.
For critical systems at the edge of chaos, where a CLT is not
valid anymore due to strong correlations, we have shown that the
relevant limit distributions appear to be $q$-Gaussians with
$q\approx 1.75$. This result is universal and independent of the
order of the maximum of the map under consideration. Our results
represent a kind of power-law generalization of the CLT, which is
relevant for entire classes of dynamical systems.
An analytical study of the present results at the edge of
chaos and of more general critical dynamical systems would be very
welcome. This might, in particular, enlighten the deep reasons
for the frequent occurrence of $q$-Gaussians in natural,
artificial and social complex systems.

This work has been supported by TUBITAK (Turkish Agency) under the Research Project number 104T148.
C.B. acknowledges financial support by EPSRC.
C.T. acknowledges partial financial support from Pronex, CNPq and Faperj (Brazilian Agencies).

\end{document}